\documentclass[prl,aps,amsmath,amssymb,floatfix,twocolumn,longbibliography]{revtex4-1}

\pdfoutput=1
\usepackage{hyperref}
\usepackage{graphicx}
\usepackage[usenames]{color}
\usepackage{bm}

\definecolor{blue}{rgb}{0,0,0}
\newcommand{\bb}[1]{\textcolor{blue}{#1}}

\newcommand*\demon{\includegraphics[width=0.2cm]{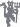}}

\def\cal#1{\mathcal{#1}}
\def\eqq#1{Eq.~(\ref{#1})}

\def\eq#1{(\ref{#1})}

\def\f#1{Fig.~\ref{#1}}

\def\c#1{~\cite{#1}}

\def\cc#1{Ref.~\cite{#1}}

\def\av#1{\langle #1 \rangle}

\def\beq{\begin{equation}}
\def\eeq{\end{equation}}
\def\bea{\begin{eqnarray}}
\def\eea{\end{eqnarray}}

\def\kt{k_{\rm B}T}

\begin{document}

\title{Demon in the machine: learning to extract work and absorb entropy from fluctuating nanosystems}
\author{Stephen Whitelam$^{\demon}$}\email{swhitelam@lbl.gov}
\affiliation{$^{\demon}$Molecular Foundry, Lawrence Berkeley National Laboratory, 1 Cyclotron Road, Berkeley, CA 94720, USA}

\begin{abstract}

We use Monte Carlo and genetic algorithms to train neural-network feedback-control protocols for simulated fluctuating nanosystems. These protocols convert the information obtained by the feedback process into heat or work, allowing the extraction of work from a colloidal particle pulled by an optical trap and the absorption of entropy by an Ising model undergoing magnetization reversal. The learning framework requires no prior knowledge of the system, depends only upon measurements that are accessible experimentally, and scales to systems of considerable complexity. It could be used in the laboratory to learn protocols for fluctuating nanosystems that convert measurement information into stored work or heat.

\end{abstract}
\maketitle

\section{Introduction}
Driving a fluctuating molecular machine or nanoscale system out of equilibrium costs energy, on average, according to the second law of thermodynamics\c{chandler1987introduction,fowler1939statistical}. For example, when averaged over many experiments, an optical trap requires an input of work to drag a colloidal particle through water, and a nanomagnetic system whose state is flipped by a magnetic field produces entropy. {\em Individual} realizations of these processes can generate work or absorb entropy, and the fluctuation relations, which generalize the second law, make statements about the distributions of work and entropy that result from various kinds of nonequilibrium processes\c{jarzynski1997nonequilibrium,evans2002fluctuation,crooks1999entropy,seifert2012stochastic}. However, if the nonequilibrium protocol involves feedback, i.e. has knowledge of the state of the system, then the {\em mean} of the work- and entropy-production distributions can be negative: on average, work can be extracted from the system, or entropy absorbed by it\c{maxwelltheory,szilard1964decrease,sagawa2012nonequilibrium,esposito2012stochastic,horowitz2014thermodynamics,parrondo2015thermodynamics,ehrich2022energetic}. Such behavior does not violate the second law because these changes are paid for by the acquisition and destruction of information, which increases the entropy of the universe. Thus an agent or {\em demon} that measures a system in order to control it can convert measurement information into work or heat. Such ``information engines'' have been demonstrated experimentally, using ratchet-like mechanisms to extract work from colloidal particles in electric or gravitational fields\c{toyabe2010experimental,saha2022bayesian}.

Here we demonstrate a technically simple and generally applicable procedure for learning feedback-control protocols for fluctuating nanosystems. We consider two model computational systems, a particle pulled by an optical trap through a viscous medium\c{schmiedl2007optimal} and an Ising model undergoing magnetization reversal\c{rotskoff2015optimal,gingrich2016near}. We introduce a demon, a deep neural network whose inputs are the elapsed time of the experiment and any information we provide it from the system, and whose outputs indicate the new values of the control parameters, the trap position or the Ising model temperature and magnetic field. We train the demon using Monte Carlo\c{whitelam2022training} and genetic algorithms\c{such2017deep,whitelam2020learning} to minimize the work done by the trap or the entropy produced by the Ising model. We allow the demon no prior knowledge of what constitutes an efficient protocol. When the input to the demon is time alone, the protocols it learns reproduce the optimal-control protocols known analytically or from numerical studies\c{schmiedl2007optimal,engel2022optimal,zhong2022limited}. When the demon is also provided the force on the particle or the magnetization of the Ising model, it learns protocols that {\em extract} work from the trap and {\em absorb} entropy from the Ising model's thermal bath, so converting measurement information into alternative forms of energy.

The learning procedure described here can be applied to experiments the same way it is applied to simulations. The order parameter that the procedure minimizes is the dissipation or work produced over the {\em entire} trajectory, averaged over many trajectories, quantities that are accessible experimentally. It does not require time-resolved information about the order parameter or gradients of the order parameter along a trajectory. The number of trajectories required for meaningful learning is not prohibitively large, in the context of prior experiments. Furthermore, it is possible to apply the learning framework to protocols of considerable complexity. The neural-network demon can accommodate an arbitrary number of input neurons (information from the system) and output neurons (control parameters of the experiment), and the learning algorithms we use have been shown empirically to work with neural networks having of order a thousand inputs and of order tens of millions of parameters\c{whitelam2022training,salimans2017evolution}.
 
\section{Model of a particle pulled by an optical trap}
We consider the first problem of~\cc{schmiedl2007optimal}, a particle at position $x$ in a potential
\beq
\label{vee}
V(x,\lambda)=\frac{1}{2} \left(x-\lambda \right)^2,
\eeq 
in units such that $\kt=1$; see \f{fig0}(a). The trap center is $\lambda$. The particle undergoes the Langevin dynamics
\beq
\label{langevin}
\dot{x}=-\partial_x V(x,\lambda) + \xi(t),
\eeq
where the noise $\xi$ has zero mean and \bb{correlation function} $\av{\xi(t) \xi(t')} = 2 \delta(t-t')$. The aim is to move the trap center from an initial position $\lambda_i=0$ to a final position $\lambda_{\rm f}=5$, in finite time $t_{\rm f}$, minimizing the work averaged over many realizations of the process. If $\lambda$ is a function of time alone then the optimal (work-minimizing) protocol is the linear form $\lambda^\star(t)=\lambda_{\rm f} (t+1)/(t_{\rm f}+2)$, for $0<t<t_{\rm f}$, with jump discontinuities at the start $(t=0)$ and end $(t=t_{\rm f})$. This protocol produces mean work $\lambda_{\rm f}^2/(t_{\rm f}+2)$\c{schmiedl2007optimal}.

To develop a feedback-control protocol for this system we introduce a demon, the deep neural network shown in \f{fig0}(b).  The neural network is fully connected, with 5 hidden layers (each of width 4 apart from the last, which is of width 10) and hyperbolic tangent activations. We apply layer norm pre-activation\c{goodfellow2016deep}. Deep neural networks are convenient ways of expressing potentially high-dimensional functions, and this particular architecture can be trained faster than single-layer nets to express rapidly-varying functions\c{whitelam2022training}. The network has as many input neurons as degrees of freedom it is provided, and as many output neurons as there are control parameters of the system. If ${\bm s}$ is the vector of state information provided to the demon, and ${\bm \lambda}$ the vector of experimental control parameters, then the demon, when queried, sets these control parameters to the values
\beq
\label{nn}
{\bm \lambda}= {\bm g}_{\bm \theta}(t/t_{\rm f},{\bm s}),
\eeq
where $t/t_{\rm f}$ is scaled time (information that is always available), ${\bm g}$ is the vector function expressed by the neural network, and ${\bm \theta}$ is the vector of neural-network parameters (weights and biases). \eqq{nn} is a parameterization of the control parameters of the experiment as a function of time and (potentially) state-dependent information. The aim is to adjust the numbers ${\bm \theta}$ by training, so that the protocol enacted by the demon achieves the desired objective.

For the trap system there is one control parameter, the trap center $\lambda$, and so the demon needs only one output neuron. In the main text we consider three different sets of inputs: time alone; or time and force; or time, particle position, and trap position. The neural network needs one, two, or three input neurons in each case.

The trap position is initially $\lambda_0=\lambda_{\rm i}$ and the particle position is $x_0 \sim {\cal N}(0,1)$, reflecting thermal equilibrium in the potential \eq{vee}. At each step $k=1,\dots,\lfloor t_{\rm f}/\Delta t \rfloor$ of the simulation we choose a new trap position by consulting the demon, $\lambda_k =  g_{\bm \theta}(t_k/t_{\rm f})$, where $t_k=k \Delta t$; we calculate the work done by this change,
\beq
\label{work}
\Delta W_k = V(x_k,\lambda_k)-V(x_k,\lambda_{k-1});
\eeq
and we update the position of the particle according to the forward Euler discretization of \eq{langevin} with timestep $\Delta t=10^{-3}$. 
%\beq
%x_k=x_{k-1}+(\lambda_k-x_{k-1}) \Delta t+ \sqrt{2 \Delta t}\, \eta,
%\eeq
%where $\eta \sim {\cal N}(0,1)$. 
At the end of the trajectory, at time $t_{\rm f}$, the trap center is set to position $\lambda_{\rm f}$, and the work updated accordingly. The total work done along the trajectory, $W$, is the sum of all changes \eq{work} plus the final-time work update. The order parameter we wish to minimize is $\phi = \av{W}$, where $\av{\cdot}$ denotes the average over $M$ independent trajectories of the procedure just described. The demon starts with all parameters set to zero, ${\bm \theta} = {\bm 0}$, and so has no prior knowledge of what constitutes a good protocol. The protocol enacted by this untrained demon is a jump at time $t_{\rm f}$ from $\lambda=\lambda_{\rm i}$ to $\lambda=\lambda_{\rm f}$, which gives mean work $\lambda_{\rm f}^2/2$. 
\begin{figure*}[] 
   \centering
   \includegraphics[width=0.7\linewidth]{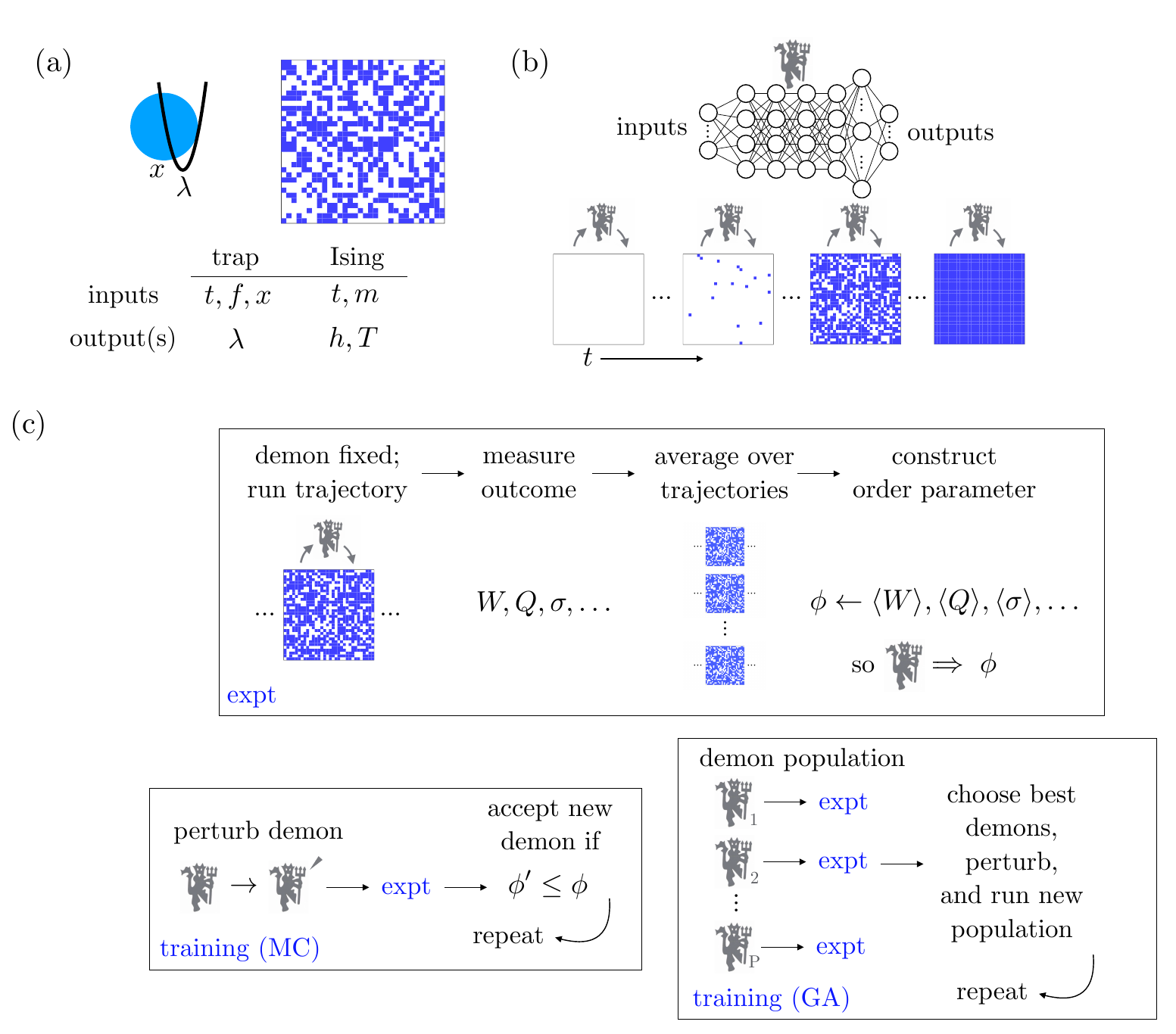} 
   \caption{(a) We develop feedback-control protocols for an overdamped particle pulled by a harmonic potential and the Ising model undergoing magnetization reversal. (b) Feedback is enacted by a demon, a deep neural network, which controls the protocol to which the system is subjected by periodically taking in information from the system and outputting new values of the system's control parameters. (c) The demon is trained to extremize a desired physical observable, such as heat or work. Dynamical trajectories of the system, generated using the demon-mandated protocol, result in an order parameter $\phi$ that is composed of ensemble averages of work, heat, entropy production, or other measurable quantities (box labeled ``expt''). The demon is trained iteratively, by Monte Carlo (box ``MC'') or genetic algorithms (box ``GA''), to extremize $\phi$. In this paper the trajectories are generated by computer simulations of model systems, but the same learning procedure could be applied to trajectories generated in laboratory experiments.}
   \label{fig0}
\end{figure*}
\begin{figure*}[] 
   \centering
   \includegraphics[width=\linewidth]{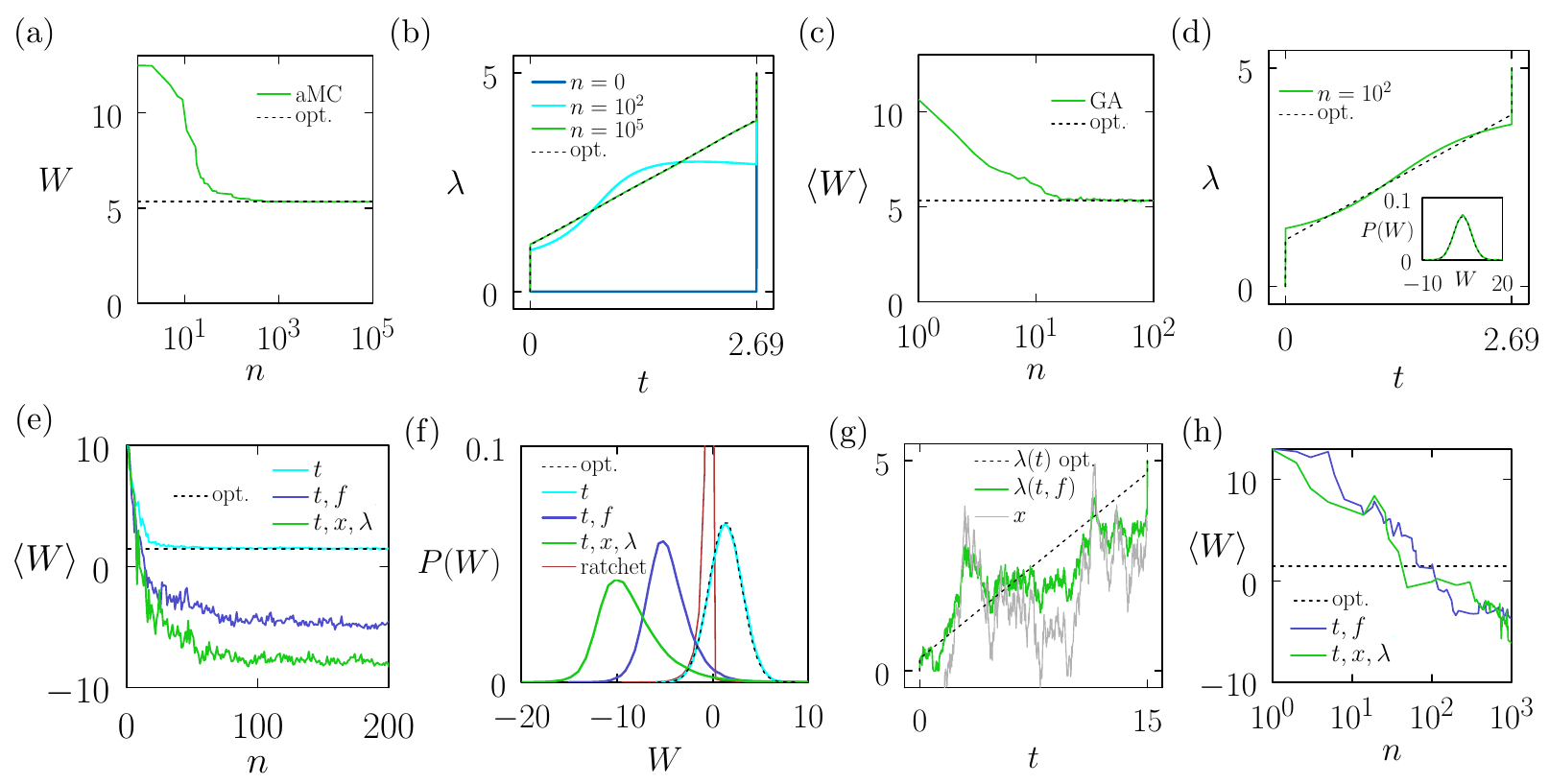} 
   \caption{Particle pulled by a trap. (a) Work $W$ as a function of aMC steps $n$ for a single trajectory of the noise-free version of \eq{langevin} (mimicking a large number of trajectories) under the neural-network demon protocol $\lambda =  g_{\bm \theta}(t/t_{\rm f})$ (green). The black dashed line is the analytic result of~\cc{schmiedl2007optimal}. (b) Time-dependent protocol $\lambda =  g_{\bm \theta}(t/t_{\rm f})$ learned after $n=0,10^2$, and $10^5$ aMC steps (colors), compared to the optimal protocol (black dashed). (c,d): as (a,b), but now the order parameter $\av{W}$ is calculated as an average over $10^4$ noisy trajectories. GA is used as a method of training; $n$ is the number of generations (here and elsewhere, the GA result indicates the performance of the best demon in each generation). The learned protocol is a near-optimal one that in practical terms cannot be distinguished from the optimal protocol: the inset to (d) shows the work distributions produced by each. (e) Mean work $\av{W}$ (averaged over $10^4$ trajectories) versus GA generation $n$ for demon protocols expressed as a function of $t$ (cyan), $t$ and $f$ (blue), and $t,x$, and $\lambda$ (green). The black dashed line is the analytic result of~\cc{schmiedl2007optimal}. (f) Work distributions for demon protocols learned after 200 generations (colors), together with that of a ratchet (red). (g) A single Monte Carlo trajectory under the demon $t,f$-protocol learned after 200 generations. (h) Similar to (e), but averaging $\av{W}$ over only $10^2$ trajectories and using aMC as a training method. Trajectory lengths: $t_{\rm f}=2.69$ (a--d) and 15 (e--h).}
   \label{fig1}
\end{figure*}

To train the demon we use the adaptive Monte Carlo (aMC) algorithm of~\cc{whitelam2022training}, or a genetic algorithm (GA)\c{GA,mitchell1998introduction,such2017deep,whitelam2020learning} using mutations only. Training is illustrated in \f{fig0}(c). aMC is an adaptive version of the Metropolis algorithm\c{metropolis1953equation}, and proceeds as follows (see box labeled ``MC''). We evaluate the order parameter $\phi$ by generating trajectories using the demon-mandated protocol (see box labeled ``expt''). We add independent Gaussian random numbers (``mutations'') to each demon parameter, and evaluate the objective $\phi'$ under the new demon. If $\phi'\leq \phi$ then we accept the new demon; otherwise, we return to the current one. This part of the procedure is zero-temperature Metropolis Monte Carlo applied to the parameters of a neural network; in addition, aMC adjusts the mean and variance of the mutations in order to propose moves that are more likely to be accepted. We use the aMC hyperparameters $\sigma_0=10^{-1}, \epsilon=10^{-2}$, and $n_{\rm scale}=10^3$, with signal norm off (we use layer norm instead)\c{whitelam2022training}. Training using a genetic algorithm (see box labeled ``GA'') proceeds by evaluating the order parameter $\phi$ under 50 randomly-initialized demons (for which each parameter is an independent Gaussian random number $\theta \sim {\cal N}(0,\sigma^2)$, with $\sigma=0.1$) and picking the 5 with the smallest values of $\phi$. The next generation of 50 demons is created by picking 49 times randomly with replacement from this set of 5 parents, and adding independent Gaussian random numbers of zero mean and variance $\sigma^2=10^{-2}$ to each parameter of each demon. The final member of the population is the unmutated best demon from the previous generation, a procedure called elitism. The order parameter is evaluated for this new set of 50 neural networks, the best 5 are chosen to be the parents of the subsequent generation, and so on. The two methods, GA and MC, are closely related: zero-temperature Metropolis MC is also a genetic algorithm with a population of size two with one parent, using elitism and mutations only. Both aMC and GA are simple to implement and have similar learning capacity to gradient-based methods\c{sexton1999beyond,rere2015simulated, such2017deep,tripathi2020rso,whitelam2021correspondence} (though they do not necessarily converge as quickly, step for step\c{whitelam2022training}). They do not require gradient information from the trajectory, making them ideally suited to laboratory experiments. GA is convenient if experiments can be run in parallel, while aMC usually requires fewer function evaluations (in this case, trajectories or experiments) to reach a prescribed value of the order parameter.

To make contact with prior results we first consider the case in which the demon knows only the elapsed time of the trajectory, and so the trap position is $\lambda =  g_{\bm \theta}(t/t_{\rm f})$. In panels (a) and (b) of \f{fig1} we consider the hypothetical limit in which the order parameter $\phi=\av{W}$ is evaluated using a large number $M \to \infty$ of trajectories (when $\lambda$ depends only on time, the mean work is a function of $\av{x}$, and this can be calculated in the limit $M \to \infty$ using a single trajectory of the noise-free version of \eq{langevin}). In panels (c) and (d) we consider the experimentally-relevant case in which a finite number ($M=10^4$) of trajectories is used. In both cases the order parameter converges to the optimal value. For $M \to \infty$ the demon learns the optimal protocol (black dashed line), while for finite $M$ it learns an approximation of it. However, the outcome of the optimal protocol cannot be distinguished from that of the learned protocol: as shown in the inset of panel (d), the work distributions produced by optimal and learned protocols are essentially identical. As $n$ increases, the learned protocol fluctuates about the optimal one, adopting a large number of different shapes that have similar outcomes; see \f{fig_profile_multi}.

The analytic optimal protocol of \cc{schmiedl2007optimal} has been reproduced by other authors using gradient descent\c{engel2022optimal} and numerical methods of optimal control\c{zhong2022limited}. These methods are powerful numerically but require information not accessible experimentally, namely gradients of the objective along the dynamical trajectory or knowledge of the deterministic Fokker-Planck equation that governs the evolution of the probability density, respectively. Here we have shown that the optimal time-dependent protocol can be learned if we know only the total work performed during a stochastic trajectory, without time-resolved input from the system, and with no prior knowledge of what constitutes an efficient protocol.

In \f{fig1}(e--h) we show that providing feedback to the demon allows it to learn protocols whose mean values of work are negative. We compare cases in which the demon is a function of time alone;  a function of time and the force $f=x-\lambda$ acting on the particle, $\lambda =g_{\bm \theta}(t/t_{\rm f},f)$; and a function of time, particle position $x$, and trap position, $\lambda=g_{\bm \theta}(t/t_{\rm f},x/\lambda_{\rm f},\lambda/\lambda_{\rm f})$. In panel (e) we show that feedback allows the mean work to be negative: the demon has learned to extract work from the system. (The demon can also extract work from the system if it is fed other information, such as the current value of the time-integrated work, but the instantaneous system coordinates are more useful in this respect; see \f{fig_supp_trap}.)

Panel (f) shows the distributions of work resulting from these protocols, together with that resulting from a simple ratchet-like mechanism\c{toyabe2010experimental} in which, at each timestep, $\lambda$ is set to $x$ if $x>\lambda$ and $\lambda \leq \lambda_{\rm f}$, and is otherwise unchanged. The ratchet can extract work from the thermal bath, but not as efficiently as the protocols learned by the demons. Panel (g) shows one trajectory of the time-force protocol learned after generation $n=200$: the demon extracts work by moving the trap toward the particle but with a bias in the direction of the final-time trap position. In panel (h) we show data similar to that of panel (e), but now taking work averages $\av{W}$ over only $10^2$ trajectories, and training using aMC. With fewer trajectories used to calculate the mean work, fluctuations of the learning process are larger, but work-extracting protocols can still be learned. Each GA generation of panel (e) requires the evaluation of $50 \times 10^4$ trajectories, while each aMC step of panel (h) requires the evaluation of $10^2$ trajectories. There exist optical traps that allow hundreds of trajectories per experiment\c{carberry2004fluctuations}, and so the number of experiments required by the demon to learn work-extracting protocols (of order 100) is not prohibitive in an experimental context.

\section{Magnetization reversal in the Ising model}
The learning framework can be applied to more complex, many-body systems in the same way, with straightforward changes made to the neural network to accommodate the required input and control parameters. Consider magnetization reversal in the Ising model, a prototype of information erasure and copying in nanomagnetic storage devices. Reducing dissipation by finding optimal time-dependent protocols for these processes\c{rotskoff2015optimal,gingrich2016near} has practical relevance for reducing computational energy demands\c{lambson2011exploring}. Here we use the learning framework to reproduce optimal time-dependent protocols that minimize dissipation upon magnetization reversal\c{engel2022optimal}. We then show that feedback control using experimentally-accessible measurements allows magnetization reversal to proceed with {\em negative} mean dissipation, taking heat from the surroundings.
\begin{figure*}[] 
   \centering
   \includegraphics[width=\linewidth]{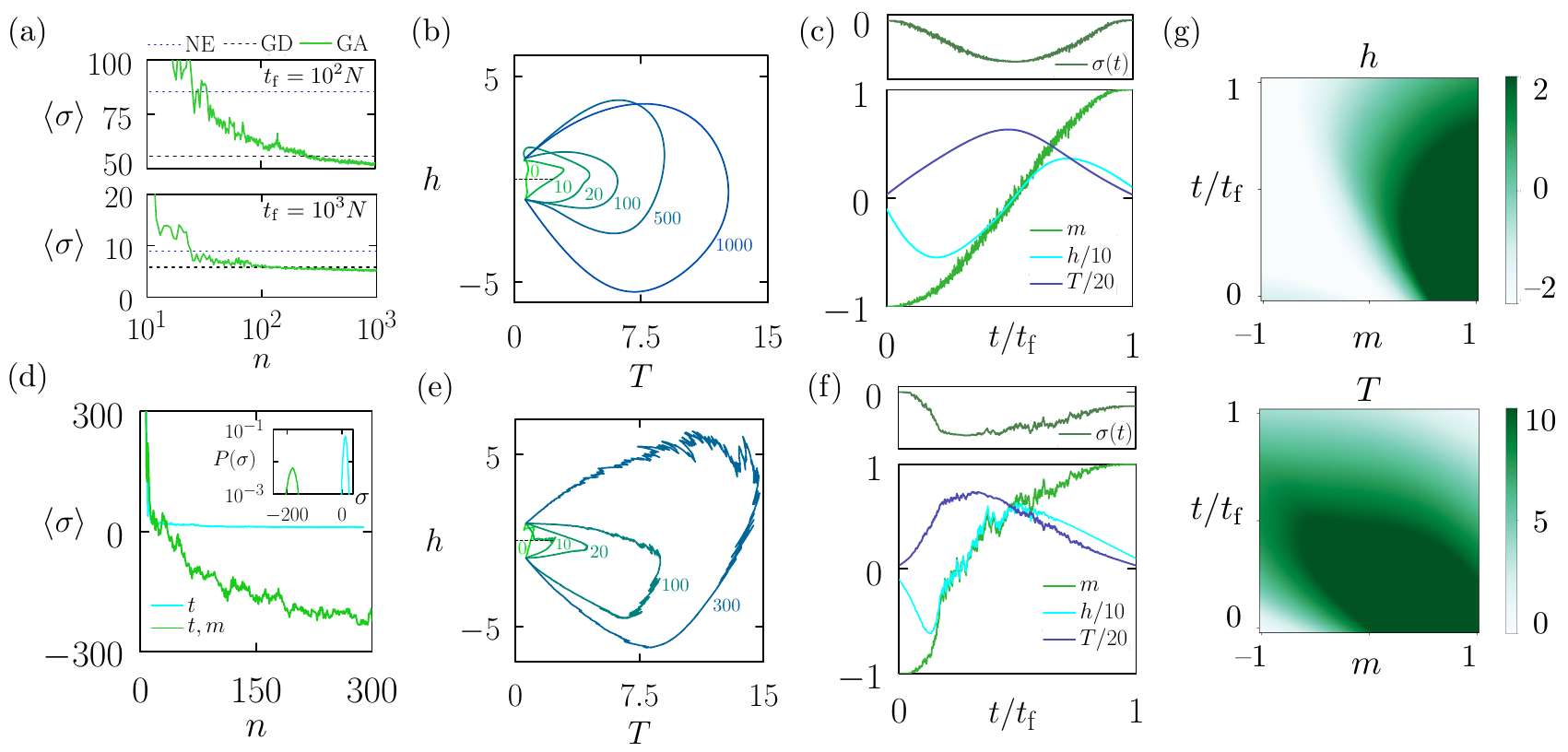} 
   \caption{Magnetization reversal in the Ising model. (a) Mean entropy production $\av{\sigma}$ under protocols learned by a demon given time as input, as a function of GA generation $n$ (green). Trajectory lengths are $10^2$ (top) or $10^3$ (bottom) Monte Carlo sweeps. Also shown are values due to entropy-minimizing protocols found by near-equilibrium approximation\c{rotskoff2015optimal} (NE) and gradient descent\c{engel2022optimal} (GD). (b) Parametric protocols $(T(t), h(t))$ learned after $n$ generations. The black dashed line is the Ising model first-order phase-transition line, which ends at the critical point\c{onsager1944crystal}. (c) A single trajectory under the protocol learned after $10^3$ generations. (d--f) Similar to (a--c), but now the demon receives time and global magnetization $m$ as inputs. The top sections of (c) and (f) show time-resolved entropy production along each trajectory (the vertical axis is $10^3$ units long): this information is not available during training, nor used by the demon. In panel (d) we also show the case in which the input is time only (cyan), and in the inset of that panel we show entropy production distributions over $10^4$ trajectories for protocols learned using time (cyan) and time and magnetization (green) as inputs. (g) Visualization of the demon feedback-control protocol learned after 300 generations. }
   \label{fig2}
\end{figure*}

We consider the 2D Ising model\c{onsager1944crystal,binney1992theory} on a square lattice of $N=32^2$ sites, with periodic boundary conditions in both directions. On each site $i$ is a binary spin $S_i = \pm 1$. The lattice possesses an energy function
\beq
\label{ising}
E = -J\sum_{\av{ij}} S_i S_j -h\sum_{i=1}^N S_i,
\eeq
in units such that $k_{\rm B} =1$. Here $J$ (which we set to 1) is the Ising coupling, and $h$ is the magnetic field. The first sum in \eq{ising} runs over all nearest-neighbor bonds, while the second runs over all lattice sites. We begin with all spins down, giving magnetization $m=N^{-1} \sum_{i=1}^N S_i=-1$. Following~\cc{engel2022optimal}, the aim is to change temperature $T$ and field $h$ from the values ${\bm \lambda}_{\rm i} = (T_{\rm i},h_{\rm i})=(0.65,-1)$ to the values ${\bm \lambda}_{\rm f} =(T_{\rm f},h_{\rm f})=(0.65,1)$, in finite time $t_{\rm f}$, ensuring magnetization reversal with minimal dissipation.  

We simulate the model using Glauber Monte Carlo dynamics. At each step of the algorithm a lattice site $i$ is chosen, and a change $S_i \to -S_i$ proposed. In \f{fig2}(a--c) we choose lattice sites deterministically, moving through all odd-numbered spins and then all even-numbered spins, in order to make contact with~\cc{engel2022optimal}. In \f{fig2}(d--g) we consider random choice of lattice site. The outcomes of these two procedures are qualitatively similar, and generate numerical values of entropy production that differ by a factor of about 2. The proposed change is accepted with the Glauber probability $\left( 1+ \exp(\beta \Delta E) \right)^{-1}$, where $\Delta E$ is the energy change under the proposed move, and $\beta=1/T$ is the reciprocal temperature. If the move is rejected, the original spin state is adopted. 

The entropy produced over the course of a simulation is 
\beq
\label{ep}
\sigma=\beta_{\rm f} E_{\rm f} -\beta_{\rm i} E_{\rm i} - \sum_{k} \beta_k \Delta E_k,
\eeq
where $E_{\rm f}$ and $E_{\rm i}$ are the final and initial energies of the system, and $\Delta E_k$ and $\beta_k$ are the energy change and reciprocal temperature at step $k$ of the simulation. This expression can be derived by considering the path probabilities of forward and reverse trajectories, or by noting that heat exchange with the bath is given by change of energy at fixed control parameters\c{seifert2012stochastic,crooks1999entropy}. The first two terms on the right-hand side of \eq{ep} cancel for any path that connects end-points of equal temperature and opposite field, and for which the magnetization reverses. 

The demon \eq{nn} now has two output neurons in order to specify the control-parameter vector ${\bm \lambda} = (T,h)$. The output layer contains a shear transformation in order to ensure that the protocol starts and ends at the required points: \bb{if $\tilde{{\bm g}}_{\bm \theta}(t/t_{\rm f},m)$ is the output of the neural network prior to the shear, then the control-parameter vector is set to}
\bea
\label{shear}
{\bm g}_{\bm \theta}(t/t_{\rm f},m) &=&\tilde{{\bm g}}_{\bm \theta}(t/t_{\rm f},m)+ (1-t/t_{\rm f}) \left[{\bm \lambda}_{\rm i}-\tilde{{\bm g}}_{\bm \theta}(0,-1)\right] \nonumber
\\&+&(t/t_{\rm f})\left[{\bm \lambda}_{\rm f}-\tilde{{\bm g}}_{\bm \theta}(1,1)\right]
\eea 
\bb{after the shear~\footnote{We use an expression similar to \eq{shear} if time alone is used as input to the neural network, with ${\bm g}_{\bm \theta}(t/t_{\rm f},m)$ replaced by ${\bm g}_{\bm \theta}(t/t_{\rm f})$.}. In addition, if $T$ resulting from \eq{shear} is less than $10^{-3}$, then it is set equal to $10^{-3}$}. Initially the demon parameters ${\bm \theta}$ are set to zero, and the protocol produced by this untrained demon jumps abruptly between initial and final values. The entropy produced by a sudden change of control parameters between initial and final values, without change of magnetization, is $2 h \beta_{\rm f} N \approx 3151$. 

The order parameter we wish to minimize is $\phi=|\av{m_{\rm f}}-1|+k_\sigma \av{\sigma}$, where $\av{\cdot}$ denotes the average over $M=10^3$ independent trajectories, $k_\sigma=10^{-4}$, $\sigma$ is given by \eqq{ep}, and $m_{\rm f}$ is the magnetization at the end of the trajectory. This order parameter is minimized if $\av{m_{\rm f}}=1$ and $\av{\sigma}$ is as small as possible. The choice $k_\sigma \ll 1$ ensures that the second term in $\phi$ is generally smaller than the first, enforcing that the primary goal of the procedure is to reverse magnetization. Once this has been achieved, the order parameter further rewards protocols that do so with as little dissipation as possible.

We start by providing the demon with time alone as input, i.e. $(T,h) = {\bm g}_{\bm \theta}(t/t_{\rm f})$. The demon acts at 1000 evenly-spaced time intervals, choosing new values of $T$ and $h$ each time it acts. We train the demon by GA (aMC results are shown in \f{fig_ga_amc}). In \f{fig2}(a) we show the mean entropy produced by learned protocols as a function of GA generation $n$ (green). Trajectory lengths are $t_{\rm f} =10^2$ or $10^3$ Monte Carlo sweeps (steps per lattice site). We also show the entropy produced by the optimal protocol obtained by the near-equilibrium approximation of \cc{rotskoff2015optimal} (blue dashed), and that found numerically by gradient descent in~\cc{engel2022optimal} (black dashed). The GA result is consistent with the latter, confirming for this problem that gradient-free and gradient-based methods have similar capacity for learning. 

In \f{fig2}(b) we show parametric protocols obtained after different evolutionary generations (here and subsequently we consider trajectories of length $t_{\rm f}=10^3$ Monte Carlo sweeps). The demon learns to avoid the large dissipation associated with the first-order phase transition and the critical point\c{rotskoff2015optimal}. Consistent with~\cc{gingrich2016near}, protocols with substantially different values of $T$ and $h$ can have similar values of dissipation (e.g. compare the values of entropy production [panel (a), bottom] due to the protocols obtained after 500 and 1000 generations [panel (b)]). In panel (c) we show a single trajectory under a protocol learned after $10^3$ generations. Spin flips against the direction of the magnetic field tend to absorb entropy, while those in direction of the field tend to produce it, and the demon controls $T$ and $h$ in order to produce as little excess entropy as possible. 

Panels (d--f) of \f{fig2} are similar to panels (a--c), but now the demon knows the elapsed time of the simulation and the global magnetization, and so $(T,h) = {\bm g}_{\bm \theta}(t/t_{\rm f},m)$. Again the demon acts at 1000 evenly-spaced times within a trajectory. Panel (d) shows that the demon learns a protocol for which the mean entropy dissipation is negative: the system has {\em absorbed} heat from the surroundings. Panels (e) and (f) show that the learned feedback-control protocols resemble those of the time-dependent protocols of panels (b) and (c), but now the demon can make small changes to control parameters in response to fluctuations of magnetization. The demon has learned to manipulate the field in response to fluctuations so that the entropy stored as spins flip against the field exceeds the entropy produced by spins flipping with the field. (We have used entropy production as an order parameter in order to make contact with prior work; we have also verified (see \f{fig_supp_ising}) that the demon can learn to do magnetization reversal with negative heat absorption, heat transfer being a more conveniently accessible quantity in experiment.)

In \f{fig2}(g) we summarize the protocol learned by the demon after 300 generations, which specifies $T$ and $h$ when given $t/t_{\rm f}$ and $m$. Neural networks are convenient ways of learning smooth (though potentially rapidly varying) protocols that interpolate to values of inputs not seen during training. In this case the deep neural net contains only 144 parameters, encoding in an efficient way a protocol that reverses magnetization with net negative dissipation.

\section{Conclusions}
We have developed feedback-control protocols for simulated fluctuating nanosystems, using genetic algorithms and Monte Carlo algorithms applied to a deep-neural-network demon. When provided with time alone, the demon learns the optimal-control protocols known analytically or from numerical studies\c{schmiedl2007optimal,engel2022optimal,zhong2022limited}. When also given feedback from the system, the demon learns protocols that extract work or store heat.

The problem considered here is one of process control\c{seborg2016process,langton2006stability}, and in machine-learning terms the approach is a form of deep reinforcement learning: training an agent, a deep neural network, to carry out time- and state-dependent actions in order to maximize a time-dependent order parameter. Monte Carlo and genetic algorithms applied to neural networks are forms of {\em neuroevolution}\c{floreano2008neuroevolution}, and are not traditionally considered part of the canon of reinforcement-learning algorithms\c{sutton2018reinforcement}. However, they are capable of solving reinforcement-learning problems, and for the problems considered here, maximizing a long-time return can be achieved even if no short-time reward is known (other forms of control algorithms have also been used to treat molecular-scale dynamical systems\c{kaiser2021data,miskin2016turning,goodrich2021designing,suen2022feedback}. 

 There exist numerical methods for controlling fluctuating nanosystems: for instance, optimal or near-optimal time-dependent protocols can be obtained by Monte Carlo path-sampling methods\c{gingrich2016near}, or by gradient-based methods used to train parameterized functions\c{engel2022optimal}. Feedback-control protocols for flashing Brownian ratchets have been developed using gradient-based methods of reinforcement learning applied to a deep neural network\c{kim2021deep}. All are powerful numerical methods but are not immediately applicable to experiment: \cc{gingrich2016near} uses Monte Carlo moves to generate the stochastic trajectory (rather than applying Monte Carlo moves only to the protocol, with trajectories generated independently of the method), while Refs.\c{engel2022optimal} and \c{kim2021deep} use gradient information not directly accessible in experiments. The significance of the present method is that it uses only experimentally-accessible data, such as the total work or heat produced by a set of stochastic trajectories. From this information alone it is possible to learn an optimal time-dependent protocol. Further, given experimentally-accessible time-resolved information, such as the force on an optical trap or the global magnetization of a nanomagnetic system, the demon can learn a feedback-control protocol to extract work or store heat. The human specifies the order parameter -- minimize work done or minimize dissipation while executing magnetization reversal -- but the demon learns autonomously, without human intervention.
 
 The key step in applying the procedure outlined in \f{fig0}(c) to experiment is to connect the neural-network demon to the outputs of the system and to the experimental apparatus that controls the protocol: the demon must take in information from the system and output new values for the protocol control parameters. A natural choice for the neural network is to have one input neuron for each piece of experimental information (e.g. time, magnetization) and one for each control parameter of the protocol (e.g. temperature, magnetic field). There are then many possible internal neural-network structures that can express the latter as a function of the former; the fully-connected deep net used here is a simple and convenient choice. (Note that if the protocol is specified as a function time alone then it is deterministic, and can be provided to the apparatus prior to the experiment in the form of a table of control-parameter values at different times.). In this paper we have fed the demon accurate information, but experiments contain imperfections. Learning can proceed in the presence of imperfection, such as feedback delay (see \f{fig_delay}), and for any real system the impact of these imperfections on learning must be assessed on a case-by-case basis.
 
The learning framework can be used in principle to control protocols of considerable complexity. The neural-network encoding of a protocol is an efficient way to cope with a large number of input and control parameters: the number of neural-net parameters scales linearly with input and output parameters, and the methods used here have been used to train neural networks containing millions of parameters\c{whitelam2022training,salimans2017evolution}. 
 
The results presented here show that work-extracting and heat-storing protocols can be learned for fluctuating nanosystems using experimentally available measurements and no prior knowledge of protocol. In the context of interacting, many-body systems such as nanomagnetic devices, these results suggest the possibility of not simply minimizing dissipation during computing, but of converting part of the entropy increase of the demon as it measures the system into a form of nanoscopic cooling as computation occurs. 

{\em Acknowledgments.---} Code for training the neural-network demon can be found at~\cc{demon_github}. I thank Megan Engel for providing the data labeled NE and GD in \f{fig2}(a), and Corneel Casert and Isaac Tamblyn for discussions. This work was performed at the Molecular Foundry at Lawrence Berkeley National Laboratory, supported by the Office of Basic Energy Sciences of the U.S. Department of Energy under Contract No. DE-AC02--05CH11231.

%\bibliography{bib}


\appendix
% \clearpage
\onecolumngrid
\section{Appendix: Supplementary Figures}
\begin{figure*}[b] 
   \centering
   \includegraphics[width=\linewidth]{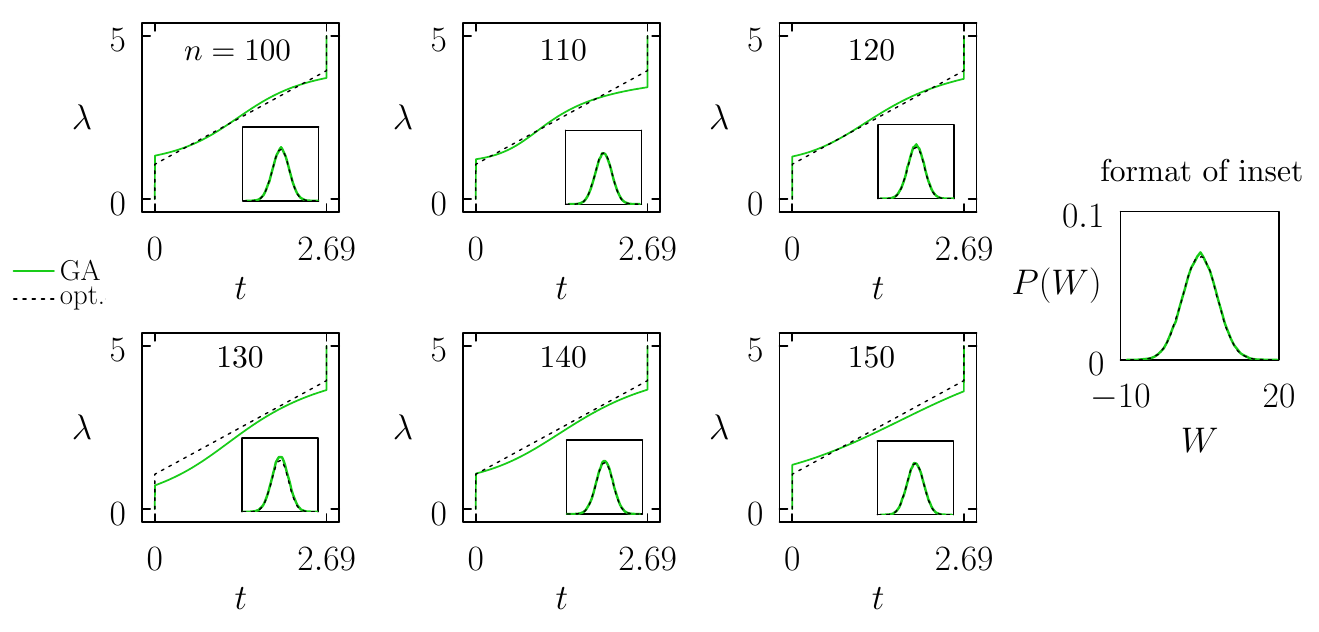} 
   \caption{Supplement to \f{fig1}. (a) As \f{fig1}(d), but with results for 5 additional generations. The shape of the learned protocol fluctuates around the optimal one, but all are similarly efficient.}
   \label{fig_profile_multi}
\end{figure*}

% \clearpage
\begin{figure*}[] 
   \centering
   \includegraphics[width=\linewidth]{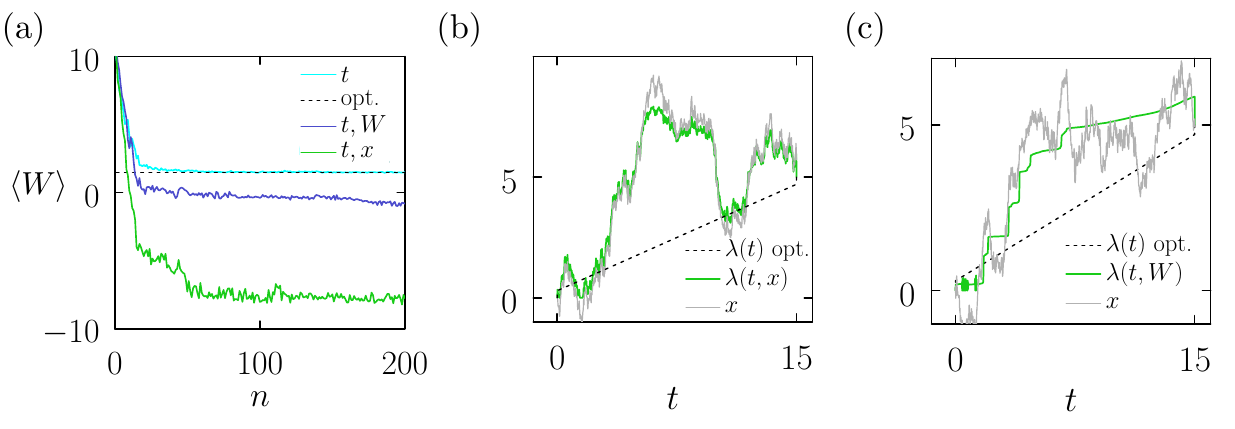} 
   \caption{Supplement to \f{fig1}: similar to \f{fig1}(e), but including the outcome of training when the demon inputs are time $t$ and time-integrated work $W$ (blue). Some work can be extracted in this case, but less than with time and bead position as input (green). Panels (b) and (c) are similar to \f{fig1}(g). (b) Knowing the bead position, the demon can move the trap so as to extract large amounts of work from position fluctuations. (c) The time-integrated work contains less information about position fluctuations, and this becomes increasingly true as time elapses. The demon can use this information to extract work, but less efficiently than if presented with the instantaneous position.}
   \label{fig_supp_trap}
\end{figure*}
% \clearpage
\begin{figure*}[] 
   \centering
   \includegraphics[width=0.75\linewidth]{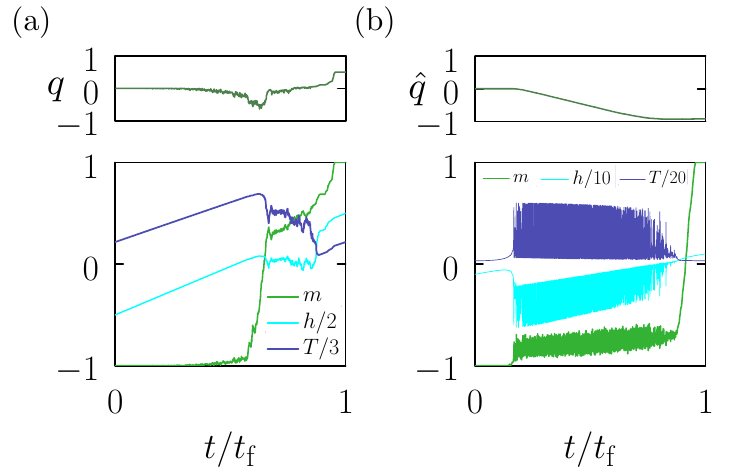} 
   \caption{Supplement to \f{fig2}: similar to \f{fig2}(f), but with heat $Q=\sum_{k} \Delta E_k$ replacing \eqq{ep} as the order parameter to be minimized. Panels (a) and (b) show magnetization-reversal cycles accompanied by heat flow out of and into the system, respectively. Here  $q\equiv Q/N$ (recall that $N=32^2$), and $\hat{q}\equiv q/200$.}
   \label{fig_supp_ising}
\end{figure*}
% \clearpage

\begin{figure*}[] 
   \centering
   \includegraphics[width=0.75\linewidth]{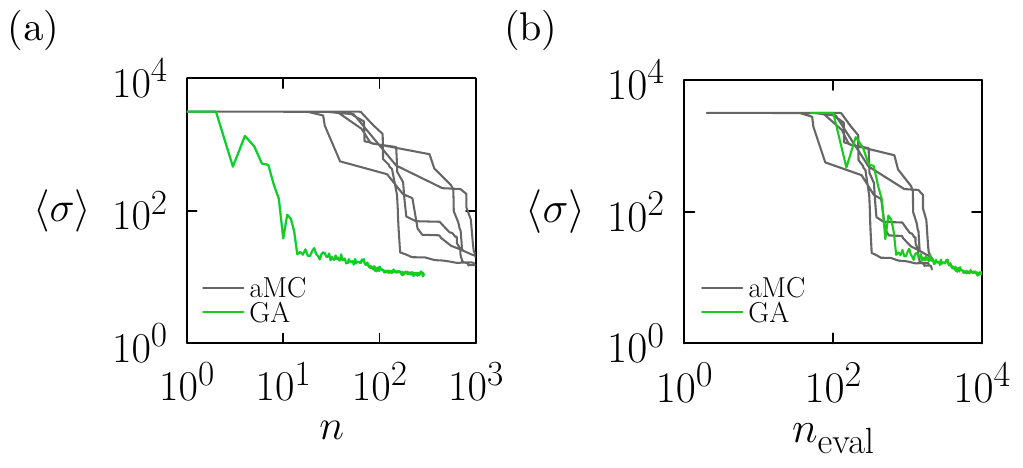} 
   \caption{Supplement to \f{fig2}: similar to \f{fig2}(d) (with the demon trained using time alone), showing the GA result of that figure (green) against 5 results obtained using aMC as a method of training (gray). The horizontal axes of panels (a) and (b) denote number of epochs (training cycles) and number of evaluations of the loss function (proportional to the number of trajectories run), respectively. GA is particularly convenient if trajectories can be run in batches or in parallel. In this example, aMC and GA converge similarly quickly per trajectory.}
   \label{fig_ga_amc}
\end{figure*}
% \clearpage
\begin{figure*}[] 
   \centering
   \includegraphics[width=0.5\linewidth]{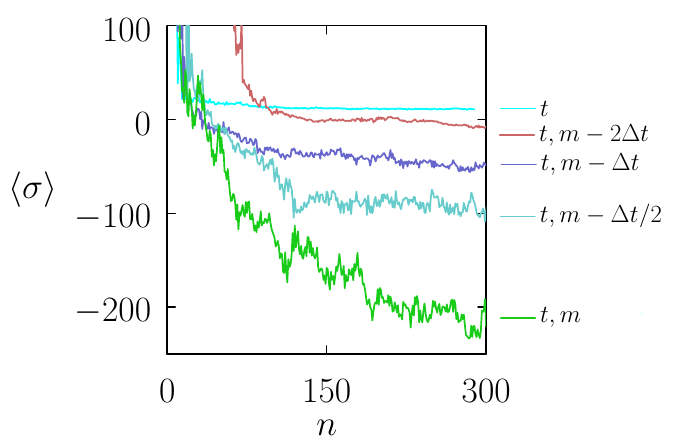} 
   \caption{Supplement to \f{fig2}: as \f{fig2}(d), but including results (red, purple, cyan) in which the demon is fed time $t$ and magnetization at a prior time, where $\Delta t = t_0/10^3$. This time delay is designed to mimic feedback delay in an experimental system. With delayed feedback the demon can still arrange for entropy to be absorbed during magnetization reversal, but less successfully than when it is fed magnetization without delay. In the presence of delay, its ability to exploit magnetization fluctuations is impaired, the more so as the delay increases.}
   \label{fig_delay}
\end{figure*}

\end{document}